\pdfoutput=1
\documentclass[10pt,letterpaper]{article}
\usepackage[top=0.85in,left=2.75in,footskip=0.75in]{geometry}
\usepackage{amsmath,amssymb}
\usepackage{changepage}
\usepackage{textcomp,marvosym}
\usepackage{cite}
\usepackage{nameref,hyperref}
\hypersetup{hidelinks}
\usepackage[nopatch=eqnum]{microtype}
\DisableLigatures[f]{encoding = *, family = * }
\usepackage[table]{xcolor}
\usepackage{array}
\usepackage{booktabs}
\usepackage{longtable}
\usepackage{pdflscape}
\usepackage{listings}
\usepackage[aboveskip=1pt,labelfont=bf,labelsep=period,
  justification=justified,singlelinecheck=off]{caption}
\usepackage{lastpage,fancyhdr,graphicx}
\usepackage{epstopdf}

\newcolumntype{+}{!{\vrule width 2pt}}
\newlength\savedwidth

\makeatletter
\renewcommand{\@biblabel}[1]{\quad#1.}
\makeatother

\fancyheadoffset[L]{2.25in}
\fancyfootoffset[L]{2.25in}
\lstdefinelanguage{Julia}{
  morekeywords={using,println},
  sensitive=true,
  morecomment=[l]{\#},
  morestring=[b]"
}
\lstdefinestyle{julia}{
  language=Julia,
  basicstyle=\ttfamily\footnotesize,
  numbers=left,
  numberstyle=\tiny,
  numbersep=7pt,
  frame=single,
  breaklines=true,
  showstringspaces=false,
  columns=fullflexible,
  keepspaces=true
}
\lstdefinestyle{prompt}{
  basicstyle=\ttfamily\footnotesize,
  frame=single,
  breaklines=true,
  showstringspaces=false,
  columns=fullflexible,
  keepspaces=true
}

\usepackage{float}
\begin{document}
\vspace*{0.2in}

\begin{flushleft}
{\Large
\textbf{Benchmarking large language models for parameter-level structural identifiability classification in epidemic ODE models.}
}
\newline
\\
Geunsoo Jang\textsuperscript{1*},
K. Sel\c{c}uk Candan\textsuperscript{1},
Gerardo Chowell\textsuperscript{2}
\\
\bigskip
\textbf{1} School of Computing and Augmented Intelligence, Arizona State University, Tempe, AZ, USA
\\
\textbf{2} Department of Population Health Sciences, School of Public Health,
Georgia State University, Atlanta, GA, USA
\\
\bigskip
* gjang12@asu.edu (GJ)
\end{flushleft}

\section*{Abstract}
Structural identifiability determines whether unknown parameters can be uniquely recovered from ideal model outputs, making it a prerequisite for interpreting parameter estimates. We evaluated whether large language models (LLMs) can reliably provide preliminary structural identifiability classifications for epidemic models. We constructed a retrospective benchmark of 88 scenarios from 14 of the 27 studies considered, encompassing 696 parameter-level formal reference classifications. Scenarios were included when sufficiently resolved structural identifiability classifications were available from published results or obtained through Julia-based symbolic analysis. Scenarios were evaluated using Domain-Aware and canonically relabeled (“Domain-Anonymized”) prompts, with ten separately submitted runs per configuration and prompt condition. Six configurations completed all scheduled responses, yielding 10,560 responses and 83,520 parameter-level decisions for the primary comparison. Overall, GPT-5.5 Instant and GPT-5.5 Thinking Medium had the highest accuracies among complete configurations (88.71\% and 88.07\%, respectively). Canonical relabeling improved both configurations and attenuated their performance decline in scenarios with many unknown parameters or state variables, although effects differed across model families. Locally identifiable parameters were the most difficult class, with the highest strict recall of 42.00\% among complete configurations; the small local reference set warrants caution. Multivariable analysis demonstrated that each additional unknown parameter reduced the odds of correct classification (odds ratio: 0.881; 95\% confidence interval: 0.852--0.910), whereas additional observed outputs increased these odds (odds ratio: 1.666; 95\% confidence interval: 1.503--1.846). Exploratory evaluations indicated strong performance for GPT-5.5 Thinking High and Claude Opus 5, despite incomplete coverage. LLMs may generate preliminary classification hypotheses or flag cases for formal review, provided their outputs are checked against published analytical results or formal symbolic methods. They should not be used to certify the structural identifiability of a complete model.

\section*{Author summary}
Epidemic models contain unknown quantities that describe disease transmission, progression, and recovery. Before estimating these parameters from data, researchers must determine whether the available observations can uniquely identify them. This property, termed structural identifiability, typically requires specialized mathematical analysis. An epidemic model can fit the observed data well even when different parameter values produce the same observed outputs. A good fit therefore does not establish that the data uniquely determine the underlying transmission or disease-progression parameters. We assembled 88 model scenarios from 14 studies, comparing LLM classifications against published or symbolic results. Moreover, we replaced disease-specific state and parameter names with neutral symbols to assess sensitivity to symbolic representation. Among the fully evaluated configurations, the GPT-5.5 configuration showed the highest agreement with formal references. Anonymization generally improved performance, particularly for complex scenarios. Overall, accuracy decreased as the number of unknown parameters increased, but improved when more outputs were observed. Our findings suggest that LLMs may help generate preliminary hypotheses or identify cases needing formal review, provided every classification is checked using published analytical results or established symbolic tools. LLM outputs should not be used to certify the structural identifiability of an entire model.

\clearpage
\newgeometry{top=0.85in,left=1in,right=1in,footskip=0.75in}

\section*{Introduction}
\label{sec:introduction}
Ordinary differential equation (ODE) models are extensively employed to simulate epidemic transmission, disease progression, vaccination effects, hospitalization rates, and mortality. These models are typically calibrated using observed data---such as incidence, prevalence, hospitalizations, or deaths--- and their estimated parameters help researchers interpret transmission mechanisms, evaluate intervention effects, and generate forecasts \cite{eisenberg2013cholera,massonis2021covid}. However, achieving a close fit between model outputs and observed data does not guarantee that the estimated parameter values are uniquely determinable. Different parameter values can sometimes generate exactly the same observable behavior, even when the outputs are observed continuously and without measurement error. Consequently, falsely declaring a parameter identifiable can support mechanistic or intervention interpretations that the specified observations do not uniquely justify.

Structural identifiability provides a rigorous theoretical framework to determine whether unknown parameters can be uniquely recovered from ideal model outputs. A globally identifiable parameter has a single, unique admissible value, whereas a locally identifiable parameter is restricted to finitely many admissible values. A non-identifiable parameter remains compatible with infinitely many values \cite{villaverde2016structural,dong2023structuralidentifiability,liyanage2026tutorial}. This distinction is crucial, as local identifiability denotes a finite degree of ambiguity rather than the absolute certainty of global uniqueness or unrestricted ambiguity non-identifiability. Furthermore, structural identifiability must be distinguished from practical identifiability, which addresses the precision of parameter estimation under finite, noisy, or sparsely sampled data. Therefore, establishing structural identifiability is an essential prerequisite before deriving epidemiological interpretations from parameter estimates based on real-world data.

Crucially, identifiability is a property of the complete model and its associated observation specifications, rather than an inherent feature of the state equations alone. Changing the observed outputs, known inputs, set of unknown parameters, or initial-condition assumptions can fundamentally alter a model's identifiability. Observing disease incidence may yield different identifiability classifications than observing prevalence or cumulative deaths. Similarly, treating previously unknown initial conditions as known can resolve parameter ambiguity, whereas treating previously known initial conditions as unknown can introduce ambiguity. Previous studies on epidemic models have extensively documented this sensitivity to data types and observational designs \cite{dankwa2022datatype,massonis2021covid,eisenberg2013cholera}. Consequently, identifiability must be evaluated independently for every unique combination of model-,observation-, and initial-condition specifications.

Formal structural identifiability analysis is traditionally executed using symbolic, differential-algebraic, or power-series-based methods. Recent software advancements have democratized access to these analyses for ODE models and can provide mathematically grounded classifications \cite{dong2023structuralidentifiability,reybarreiro2023benchmarking,liyanage2026tutorial}. However, each model must be encoded, requiring precise specification of state equations, observed outputs, unknown parameters, known quantities, external inputs, and initial conditions. Because even a minor change in any of these components can alter the underlying formal mathematical problem and its corresponding solution, preparing and analyzing multiple related model variants requires substantial computational and manual effort.

Recently, large language models (LLMs) have emerged as potential tools to support preliminary reasoning concerning structural identifiability. LLMs can process model equations and accompanying assumptions directly from a text prompt, rapidly generating parameter-level classification hypotheses that may help prioritize models or parameters for formal review without requiring users to interact with specialized symbolic software. However, these classifications are not mathematical proofs and should not be treated as certificates of identifiability; they must be evaluated against published analytical results or formal symbolic analyses. Broader scientific-reasoning benchmarks, such as SciBench, have similarly highlighted the necessity of evaluating LLMs on domain-specific, college-level scientific problems rather than relying exclusively on general-purpose benchmarks \cite{wang2024scibench}. Structural identifiability provides a particularly stringent setting because the answer depends on the complete model–output–unknown-set–initial-condition specification, and closely related model structures can yield different classifications under different observation designs \cite{villaverde2016structural,chowell2023structural}.

LLM performance in this specialized setting likely depends on both prompt representation and overall scenario complexity. Model-specific state and parameter names may inadvertently provide semantic cues that are absent when the identical equations are expressed using neutral, generalized symbols. To address this, we compared Domain-Aware and canonically relabeled (“Domain-Anonymized”) prompts to evaluate sensitivity to symbolic representation while preserving the model equations and assumptions. Because the relabeling retained equation topology and the complete observation specification, this comparison cannot isolate terminology from tokenization, notational regularity, indexing effects, or possible familiarity with the underlying equations. We also evaluated performance across various provider-defined model configurations and reasoning settings. Scenario complexity was quantified by analyzing the numbers of state variables and unknown parameters, while the number of observed outputs served as a measure of available observation information.

We assembled a retrospective benchmark of 88 model--observation--initial-condition scenarios from 14 source studies to evaluate LLM-based structural identifiability classification. Our primary objective was to quantify the agreement between parameter-level LLM classifications and formal reference classifications under matched model, observed-output, unknown-parameter, and initial-condition specifications. Our secondary objectives were to assess the effects of canonical symbolic relabeling, compare performance across formal identifiability classes and LLM configurations, examine associations with scenario complexity and observation settings, and distinguish repeated-run consistency from correctness. The primary analyses included the six configurations with complete coverage across both prompt conditions and all ten scheduled runs. Configurations with incomplete response coverage were evaluated separately in exploratory available-case analyses.

\section*{Materials and Methods}

\subsection*{Structural identifiability framework}

Consider a parameterized epidemic model defined by the following system:
\[\dot{\mathbf{x}}(t) = \mathbf{f}\bigl(\mathbf{x}(t),\boldsymbol{\theta},\mathbf{u}(t)\bigr), \qquad \mathbf{x}(0)=\mathbf{x}_0, \qquad \mathbf{y}(t) = \mathbf{h}\bigl(\mathbf{x}(t),\boldsymbol{\theta},\mathbf{u}(t)\bigr),\]
where $\mathbf{x}(t)$ represents the state vector, $\boldsymbol{\theta}$ contains the unknown parameters, $\mathbf{u}(t)$ denotes known external inputs, and $\mathbf{y}(t)$ is the observed output vector. The function $\mathbf{f}$ dictates the underlying system dynamics, whereas $\mathbf{h}$ maps the internal model states, parameters, and known inputs to the observed outputs. The initial conditions $\mathbf{x}_0$ were specified as known, unknown, or partially known for each benchmark scenario.

Structural identifiability dictates whether a specific unknown parameter can be uniquely resolved from continuous, noise-free observations of the model outputs over a continuous time interval. Assuming generic admissible values for all the unknown quantities, a parameter $\theta_i$ is globally structurally identifiable if the following condition holds:
\[ \mathbf{y}(t;\boldsymbol{\theta},\mathbf{x}_0) = \mathbf{y}(t;\widetilde{\boldsymbol{\theta}}, \widetilde{\mathbf{x}}_0) \quad \text{for all } t \quad \Longrightarrow \quad \theta_i=\widetilde{\theta}_i. \]
In this context, known initial conditions are held constant, whereas unknown initial conditions are permitted to vary across their respective admissible spaces. A parameter is classified as locally identifiable if the observed outputs restrict it to a finite set of admissible values without yielding a unique, single solution. Conversely, a parameter is deemed non-identifiable when an infinite number of admissible values remain compatible with the same model outputs \cite{villaverde2016structural,dong2023structuralidentifiability,liyanage2026tutorial}.

It is important to emphasize structural identifiability focuses purely on theoretical uniqueness under idealized observation conditions, distinguishing it from practical identifiability, which evaluates parameter estimation given a finite, noisy, or sparse dataset. An identifiability classification is intrinsically tied to the holistic analysis specification; this encompasses the model equations, observed outputs, unknown parameters, known inputs, and initial-condition assumptions. Consequently, identical state equations can yield radically different identifiability classifications if the observed outputs or initial-condition assumptions change. Each benchmark scenario was defined by a complete model--output--initial-condition specification.

Researchers have frequently assessed structural identifiability using specialized symbolic or differential-algebraic software. Prominent available implementations include Differential Algebra for Identifiability of Systems (DAISY) \cite{bellu2007daisy}, Generating Series for testing Structural Identifiability (GenSSI) \cite{chis2011genssi}, Structural Identifiability taken as Extended-Generalized Observability with Lie Derivatives and Decomposition (STRIKE-GOLDD) \cite{villaverde2016structural}, Structural Identifiability Analyzer (SIAN) \cite{hong2019sian}, and \texttt{StructuralIdentifiability.jl} \cite{dong2023structuralidentifiability}. These computational tools vary significantly in their algorithmic foundations, supported mathematical structures, and computational resource requirements \cite{reybarreiro2023benchmarking}. Evaluating or comparing the relative performance of these formal identifiability software tools was beyond the scope of this study.

\subsection*{Retrospective benchmark design}
We used purposive source selection to construct a retrospective benchmark of epidemic ODE models. We assessed 27 source papers for the availability of sufficiently detailed model specifications and structural identifiability classifications reported in the literature or obtainable through Julia-based symbolic analysis \cite{eisenberg2013cholera,Althaus2014EbolaR,Rivers2014EbolaInterventions,Webb2015EbolaContactTracing,Tuncer2016RiftValley,TuncerLe2018Outbreak,roosa2019computational,giordano2020sidarthe,lin2020conceptual,Lizarralde2020Dengue,ndairou2020wuhan,Tang2020UpdatedRisk,Deressa2021Ethiopia,Greenhalgh2021Generalized,massonis2021covid,dankwa2022datatype,Fosnaric2022Slovenia,Gallo2022COVIDPractical,sauer2022early,wu2022covid,Zhang2022HumidityInfluenza,bergstrom2026prior,liyanage2026tutorial,korsah2026reduction,zhang2021integrated,akman2025identifiability,cuevas2024vaccination}. From 14 of these papers, we reconstructed 101 candidate scenarios. Eight scenarios were excluded because Julia-based symbolic analysis did not yield a usable reference classification. Five additional scenarios from Dankwa et al.~\cite{dankwa2022datatype} were excluded because the available results established only that some parameters were at least locally identifiable, without resolving whether they were globally identifiable or locally but not globally identifiable. The final scored benchmark comprised 88 distinct scenarios and 14 papers (Table~\ref{tab:benchmark-characteristics} and S1 Table). A scenario was defined by the comprehensive combination of its model equations, observed outputs, unknown parameters, known inputs, and initial-condition assumptions. Consequently, the identical state equations were treated as entirely separate scenarios if any constituent analysis component differed. Each LLM prompt specified a single structural identifiability problem, rather than referring broadly to a general model framework. This selection process was designed to assemble a benchmark for comparative evaluation and did not constitute a systematic review of the literature.

To meet eligibility criteria, a scenario's equations and underlying analysis assumptions had to be unambiguously reconstructable. Furthermore, a directly comparable, formal three-class classification was required, either sourced directly from the original publication or derived via our subsequent Julia-based analysis. Scenario variants lacking a directly comparable published result---and which also yielded unavailable, failed, or incomplete Julia analyses---were excluded. For instance, scenarios were discarded if the existing literature could not definitively distinguish between globally or locally identifiable. In instances where a formal reference classified only a subset of the requested unknown parameters, our parameter-level scoring was restricted exclusively to that verified subset. Thorough documentation of scenario identifiers and reference provenance was maintained at both the scenario and individual parameter levels. Fig~\ref{fig:benchmark-workflow} visually outlines the comprehensive benchmark-selection and evaluation workflow.

\begin{figure}[htbp]
\centering
\fbox{%
\begin{minipage}{0.90\linewidth}
\centering
27 source papers considered\\
$\Downarrow$\\
14 source papers selected; 13 did not contribute scenarios\\
$\Downarrow$\\
101 candidate scenarios reconstructed\\
$\Downarrow$\\
13 scenarios excluded:\\
No usable Julia results or unresolved global/local distinction (Dankwa et al.)\\
$\Downarrow$\\
88 scenarios retained in the final benchmark\\
$\Downarrow$\\
696 parameter-level reference classifications from published results
or \texttt{StructuralIdentifiability.jl}\\
$\Downarrow$\\
Domain-Aware and Domain-Anonymized prompts\\
$\Downarrow$\\
Ten scheduled runs per scenario and prompt condition
for each LLM configuration\\
$\Downarrow$\\
Parameter-level comparison with formal reference classifications
\end{minipage}%
}
\caption{\textbf{Workflow for Benchmark construction and evaluation.}
 The quantitative analyses included only parameter classifications that possessed a directly comparable formal reference. Subsequent sections describe the evaluated LLM configurations and repetition protocol.}
\label{fig:benchmark-workflow}
\end{figure}

\begin{table}[htbp]
\centering
\footnotesize
\caption{Benchmark composition and scenario characteristics. The numerical ranges detail the minimum and maximum values across all scenarios derived from each respective source.}
\label{tab:benchmark-characteristics}
\begin{tabular}{lrrrrr}
\toprule
Source study & Scenarios & States & Unknown parameters & Outputs & Initial conditions \\
\midrule
Dankwa et al.~\cite{dankwa2022datatype} & 21 & 3--7 & 2--10 & 1--4 & Unknown \\
Bergstr{\"o}m et al.~\cite{bergstrom2026prior} & 1 & 2 & 3 & 1 & Known \\
Sauer et al.~\cite{sauer2022early} & 2 & 4--5 & 3--4 & 1 & Unknown \\
Roosa and Chowell~\cite{roosa2019computational} & 2 & 5--7 & 3--7 & 1 & Unknown \\
Liyanage et al.~\cite{liyanage2026tutorial} & 32 & 4--7 & 3--12 & 1--4 & Known; unknown \\
Wu et al.~\cite{wu2022covid} & 1 & 5 & 4 & 1 & Unknown \\
Zhang et al.~\cite{zhang2021integrated} & 7 & 12 & 5 & 1--3 & Known \\
Eisenberg et al.~\cite{eisenberg2013cholera} & 2 & 4 & 6 & 1 & Known; unknown \\
Massonis et al.~\cite{massonis2021covid} & 6 & 6--7 & 10--14 & 2--3 & Known; unknown \\
Cuevas-Maraver et al.~\cite{cuevas2024vaccination} & 3 & 10--18 & 10--19 & 1--2 & Mixed \\
Nda{\"i}rou et al.~\cite{ndairou2020wuhan} & 1 & 8 & 12 & 2 & Unknown \\
Giordano et al.~\cite{giordano2020sidarthe} & 4 & 8--9 & 16 & 5 & Known; unknown \\
Akman~\cite{akman2025identifiability} & 2 & 9 & 17--18 & 4--6 & Known; mixed \\
Korsah et al.~\cite{korsah2026reduction} & 4 & 8--10 & 20--21 & 1--10 & Unknown \\
\bottomrule
\end{tabular}
\end{table}

For every eligible scenario, the LLM generated classification was cross-referenced against a parameter-level formal reference. We prioritized results reported directly in the source literature, provided that the original model equations, observed outputs, unknown parameters, known inputs, and initial conditions matched our reconstructed scenario. In cases where the source study lacked a directly comparable result, we manually encoded the exact scenario into \texttt{StructuralIdentifiability.jl} to generate the formal reference classification \cite{dong2023structuralidentifiability,liyanage2026tutorial}. All Julia-derived classifications were executed using the \texttt{assess\_identifiability} function within Julia version 1.12.5 and \texttt{StructuralIdentifiability.jl} version 0.5.22. Any Julia analysis that failed, was interrupted, or did not complete was excluded and never treated as a valid identifiability result.

The model equations, original prompts, and scenario characteristics were retained unchanged. The final reference table contained 696 parameter classifications (S1 Table), categorized into 477 globally identifiable, 25 locally identifiable, and 194 non-identifiable parameters. In this parameter set, 277 classifications were manually extracted from the source studies, whereas the remaining 419 were computationally derived via \texttt{StructuralIdentifiability.jl}. Scenarios featuring known, unknown, or mixed initial-condition settings were treated as distinct scenarios. Similarly, alternative observed-output settings were encoded separately to ensure that each formal reference corresponded to one complete scenario specification. Crucially, specific numerical initial conditions and symbolically known generic initial conditions were not assumed to be interchangeable; each reference was matched to the initial-condition setting supplied to the LLM.

\subsection*{LLM configurations and prompt execution}

This benchmark considered LLM configurations from three model families: GPT from OpenAI \cite{openai2026gpt55}, Claude from Anthropic \cite{anthropic2026claudemodels}, and Gemini from Google \cite{google2026geminiflash,google2026geminipro}. The primary complete-coverage Application Programming Interface (API) comparison examined GPT-5.5 Instant, GPT-5.5 Thinking Medium, Claude Sonnet 4.5, Claude Opus 4.5, Gemini 3.5 Flash, and Gemini 3.1 Pro Preview. Exploratory API evaluations included GPT-5.5 Thinking High, Claude Sonnet 5, and Claude Opus 5. All API experiments were run through the OpenAI-compatible Arizona State University (ASU) Artificial Intelligence and Machine Learning (AIML) gateway \url{https://api-main.aiml.asu.edu/v1} between August 7 and 20, 2026. Exact gateway model identifiers are reported in S2 Table. The complete-coverage Opus configuration is named Claude Opus 4.5 throughout this report, matching the submitted gateway identifier and the model identifier in the archived responses (\texttt{aws/claude4\_5\_opus}); these records do not establish a provider-version-pinned backend. GPT-5.5 Thinking Medium and Thinking High used \texttt{reasoning\_effort=medium} and \texttt{reasoning\_effort=high}, respectively; all other configurations used gateway or provider defaults.

We represented each scenario using two matched prompts. The Domain-Aware prompt retained model-specific state and parameter names. In the Domain-Anonymized prompt, a deterministic one-to-one mapping replaced state-variable symbols with $x_i$, unknown-parameter symbols with $p_i$, explicitly designated constants with $c_i$, and external-input symbols with $u_i$. Initial-condition symbols and observation expressions changed accordingly because they contained the mapped state, parameter, constant, or input symbols. Generic output labels such as $y_1$ were retained. Paper and model names were file metadata and were not included in the submitted prompt. Mathematical operators, coefficients, equation topology, observed-output definitions, unknown set, known-quantity status, and initial-condition assumptions were otherwise preserved. The transformation therefore standardized symbolic representation but did not remove the mathematical structure of the model. This comparison was designed to evaluate sensitivity to notation and representation, not to isolate terminology completely or eliminate all cues that might reveal the source model.

Both prompt variations instructed the LLM to classify every unknown parameter as globally identifiable, locally identifiable, or non-identifiable. The prompts omitted any request for an inconclusive category. During API experiments, we mandated a structured JSON response encompassing four specific fields; \texttt{global\_identifiable}, \texttt{local\_identifiable}, \texttt{non\_identifiable}, and \texttt{identifiable\_combinations}. Identifiable combinations were archived separately but were not included in parameter-level accuracy calculations. Every prompt adhered to the standardized template illustrated in Fig~\ref{fig:prompt-template}. The complete equations and scenario-specific values replaced the bracketed fields before each run. To ensure impartial evaluation, the formal reference classification and original source-study conclusions were never provided to the submitted prompt.

\begin{figure}[htbp]
\centering
\begin{minipage}{0.96\linewidth}
\begin{lstlisting}[style=prompt]
For the following ODE model, determine which parameters are globally identifiable, locally identifiable, or non-identifiable from the observed output. Do not use tools or code; reason from the model equations only.

ODE system:
[Complete system of differential equations]

State variables:
[State-variable list]

Unknown parameters:
[Parameters to be classified]

Known constants or inputs:
[Known quantities and external inputs]

Initial conditions:
[Known or unknown initial-condition specification]

Observed output:
[Complete observation function or functions]

Assumptions:
[Model and analysis assumptions]

Report:
1. Classification of each parameter.
2. If individual parameters are not identifiable but identifiable combinations exist, state those combinations.
\end{lstlisting}
\end{minipage}
\caption{\textbf{Standardized LLM prompt template.} Every benchmark scenario utilized this identical model--observation--initial-condition format.}
\label{fig:prompt-template}
\end{figure}

Each request was dispatched to the gateway's endpoint and contained one user message. No system message, conversational history, tool definitions, files, web search, code execution, or memory was supplied by the investigators. In fact, the prompt explicitly forbade the use of external tools or code. We appended a formatting instruction requiring a single JSON object containing only the requested fields, with no extraneous conversational prose. No explicit sampling parameters (top-$p$, maximum-output-token setting, or seeds) were provided, meaning the model relied on default gateway or provider settings. We established a client-side timeout of 600 seconds. If a request failed, it was automatically retried up to three times during a batch execution. Persistently failed requests were marked as eligible for resubmission during subsequent completion runs, whereas successful responses were permanently logged and never overwritten. We did not manipulate prompt-cache keys or caching control parameter; thus, any provider-level implicit caching remained unexposed and uncontrolled. Each API configuration was evaluated under both prompt conditions in ten separately submitted, stateless runs. These runs repeated the 88 identical benchmark scenarios rather than sampling new ones. Therefore, a complete configuration comprised 1,760 responses ($88$ scenarios $\times$ 2 prompt conditions $\times$ 10 runs) and 13,920 parameter decisions ($696$ reference parameters $\times$ 2 $\times$ 10). Only successfully returned responses were analyzed, and response coverage was reported for each configuration. Responses were parsed automatically, and a parameter was assigned a classification only when it was placed in exactly one of the three categories. Any missing, conflicting, or unparseable prediction was categorized as inconclusive, and we explicitly refrained from any manual relabeling or data imputation.

\subsection*{Evaluation metrics and analysis}

The primary evaluation unit was the individual parameter classification. For each LLM configuration, prompt condition, and repeated run, we calculated parameter-level accuracy as the proportion of scored parameters for which the predicted class exactly matched the formal reference classification. For each complete configuration, prompt condition, and run, accuracy was calculated over 696 reference parameter classifications. To preserve granular variance, the ten repeats were analyzed separately rather than aggregated via majority voting; accuracy was reported as the mean and standard deviation across repeats. Consequently, this standard deviation reflects the response variation inherent in repeating identical queries, rather than statistical uncertainty stemming from sampling diverse models or source papers. Confusion-matrix counts were cumulatively pooled across the runs to summarize repeated decisions on the identical reference parameters.

Validating three-class parameter accuracy required exact agreement between the predicted and formal categories: globally identifiable, locally identifiable, or non-identifiable. Because the same accuracy can arise from substantively different errors, we also performed a secondary binary analysis to distinguish falsely declaring a non-identifiable parameter identifiable from incorrectly rejecting an identifiable parameter. Global and local classes were grouped as identifiable (positive), whereas non-identifiable was treated as negative. A false positive (FP) therefore represented a formally non-identifiable parameter that the LLM classified as global or local, while a false negative (FN) represented a formally global or local parameter that the LLM classified as non-identifiable. True positives (TP) and true negatives (TN) were defined analogously. We summarized these two error directions using the false-positive rate (FPR), the proportion of formally non-identifiable decisions incorrectly classified as identifiable, and the false-negative rate (FNR), the proportion of formally identifiable decisions incorrectly classified as non-identifiable:
\[ \mathrm{FPR}=\frac{\mathrm{FP}}{\mathrm{FP}+\mathrm{TN}}, \qquad \mathrm{FNR}=\frac{\mathrm{FN}}{\mathrm{FN}+\mathrm{TP}}.\]
Inconclusive predictions were counted as incorrect in the three-class accuracy calculation because they did not provide the required formal class. They were reported separately as abstentions and excluded from the binary FPR and FNR denominators rather than being counted as negative predictions. Because the binary analysis combines global and local identifiability, it does not capture errors between these two classes. We therefore calculated class-specific recall separately for globally identifiable, locally identifiable, and non-identifiable parameters. 

To assess sensitivity to canonical symbolic relabeling, we compared Domain-Aware and Domain-Anonymized prompting within each complete configuration. Effects were quantified as changes in accuracy, FP, FN, FPR, and FNR. Configurations with incomplete response coverage were analyzed separately using available responses. 

We characterized overall model complexity using three core metrics: the total numbers of state variables, unknown parameters, and observed outputs. Initial-condition settings were treated as an independent analysis condition. To facilitate descriptive subgroup analyses, we binned unknown-parameter counts into five distinct categories: 2--3, 4--5, 6--7, 8--13, and 14 or more. Similarly, state counts were grouped as 2--4, 5, 6--7, 8--10, and 11 or more, whereas observed-output counts as 1, 2, and 3 or more. We deliberately designed these ranges to prevent the formation of sparsely populated exact-count brackets; their scenario counts are provided in S3 Table. Results were stratified by prompt condition, model configuration, each complexity characteristic, and initial-condition setting (known, mixed, or unknown). A mixed setting contained both known and unknown initial quantities. These subgroup analyses were descriptive because the scenario characteristics were often correlated and unevenly distributed.

We employed multivariable logistic regression to evaluate the influence of individual scenario characteristics on the probability of correct parameter classification \cite{hosmer2013applied}. The analysis included 83,520 parameter-level predictions from the six LLM configurations with complete ten-run coverage. For parameter-level prediction $i$, let $C_i=1$ when the predicted three-class label matched the formal label and $C_i=0$ otherwise. The primary regression model was defined as
\[ \begin{aligned} \operatorname{logit}\!\left\{\Pr(C_i=1)\right\} &=\beta_0+\beta_1 S_i+\beta_2 P_i+\beta_3 O_i+\boldsymbol{\gamma}^{\mathsf T}\mathbf{Z}_i, \end{aligned} \]
where $S_i$, $P_i$, and $O_i$ denote the numbers of state variables, unknown parameters, and observed outputs, respectively. The vector $\mathbf{Z}_i$ contains initial-condition setting, prompt condition, LLM configuration, and formal identifiability class. These variables were included simultaneously, so each association was estimated while accounting for the others. The counts were entered as continuous variables, and $\exp(\beta_1)$, $\exp(\beta_2)$, and $\exp(\beta_3)$ represent the odds ratios associated with one additional state variable, unknown parameter, and observed output, respectively. Because multiple parameter predictions came from the same scenario, we used scenario-clustered robust standard errors to account for their within-scenario correlation \cite{williams2000robust}. We additionally tested whether the associations between scenario complexity and classification correctness differed among LLM configurations. We also tested whether the effect of anonymization differed among LLM configurations.

Repeated-run stability was evaluated separately for each complete configuration and prompt condition across the same 696 scenario--parameter cases. A case had a consistent valid classification when all ten submissions returned the same one of the three valid classes. We separately counted cases that were correct on all ten submissions and cases that repeatedly returned the same incorrect valid class. These summaries describe repeated submissions of a fixed benchmark rather than independently sampled scenarios.

\section*{Results}

\subsection*{Overall performance and anonymization effects}

We evaluated LLM performance using Domain-Aware and Domain-Anonymized versions of 88 benchmark scenarios, with ten runs scheduled for each full-benchmark configuration and prompt condition. Response completion and analysis status are summarized in S2 Table. Six configurations completed all 1,760 scheduled full-benchmark responses and formed the primary comparison, producing 10,560 responses and 83,520 parameter decisions in total. The exploratory configurations did not achieve complete benchmark coverage. GPT-5.5 Thinking High, Claude Sonnet 5, and Claude Opus 5 returned 1,584 (90.00\%), 1,498 (85.11\%), and 1,732 (98.41\%) of their 1,760 scheduled responses, respectively.

\begin{figure}[H]
\centering
\includegraphics[width=0.9\linewidth]{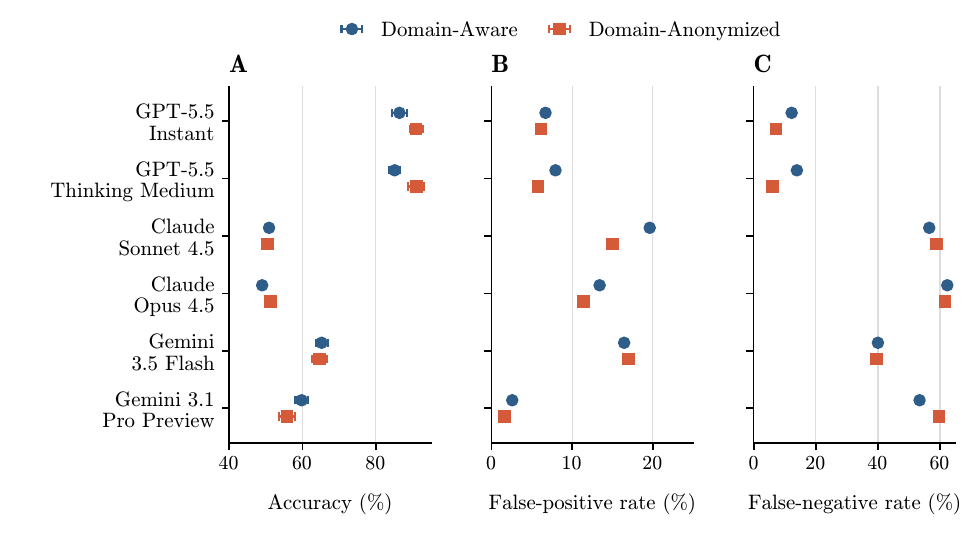}
\caption{\textbf{Overall parameter-level performance under Domain-Aware and Domain-Anonymized prompts.} (A) Three-class accuracy. (B) False-positive rate (FPR). (C) False-negative rate (FNR). Points show the mean across ten runs and error bars show the standard deviation.}
\label{fig:overall-performance}
\end{figure}

Classification accuracy varied substantially among the six configurations with complete benchmark coverage (Fig~\ref{fig:overall-performance}; S4 Table). However, little difference was observed between GPT-5.5 Instant and Thinking Medium. Under Domain-Anonymized prompting, GPT-5.5 Thinking Medium achieved an accuracy of $91.02 \pm 2.16\%$, compared with $90.99 \pm 1.80\%$ for GPT-5.5 Instant; the observed accuracies were nearly identical. When Domain-Aware and Domain-Anonymized decisions were pooled, GPT-5.5 Instant achieved an overall accuracy of 88.71\%, slightly higher than the 88.07\% achieved by GPT-5.5 Thinking Medium. Thus, GPT-5.5 Thinking Medium did not provide a clear improvement in classification accuracy over GPT-5.5 Instant in the complete-coverage comparison. The FPR measures how often formally non-identifiable parameters were incorrectly classified as identifiable, whereas the FNR measures how often formally identifiable parameters were incorrectly classified as non-identifiable. The two GPT-5.5 configurations showed the most balanced error profiles. Under Domain-Anonymized prompting, GPT-5.5 Thinking Medium had an FPR of 5.77\% and the lowest FNR among the complete configurations (6.00\%), while GPT-5.5 Instant had an FPR of 6.13\% and FNR of 7.11\%. Gemini 3.1 Pro Preview had the lowest FPR (1.65\%) but a high FNR of 59.66\%, while Claude Opus 4.5 had an FPR of 11.39\% and the highest FNR (61.67\%). Consequently, the Claude and Gemini configurations showed a greater tendency than the GPT-5.5 configurations to classify identifiable parameters as non-identifiable.

Consistency across repeated submissions was distinct from correctness (S5 Table). Under Domain-Anonymized prompting, GPT-5.5 Instant returned the same valid classification on all ten submissions for 561/696 (80.60\%) scenario--parameter cases, while GPT-5.5 Thinking Medium did so for 582/696 (83.62\%). The proportions correct on all ten submissions were lower, at 549/696 (78.88\%) and 560/696 (80.46\%), respectively. The corresponding Domain-Aware proportions were 506/696 (72.70\%) for Instant and 499/696 (71.70\%) for GPT-5.5 Thinking Medium. High consistency could also reflect persistent error. Claude Opus 4.5 returned the same valid classification on all ten submissions for 671/696 (96.41\%) cases under Domain-Anonymized prompting, but only 346/696 (49.71\%) were correct on every submission; 325/696 (46.70\%) repeatedly received the same incorrect class. This distinction shows why low variation in aggregate accuracy or repeated agreement cannot substitute for comparison with formal references.

\begin{table}[htbp]
\centering
\footnotesize
\caption{Available-case performance metrics for incomplete ten-repeat configurations. Accuracy is presented as the mean $\pm$ standard deviation across the ten runs. False-positive rate (FPR) and False-negative rate (FNR) are pooled across successfully returned responses.}
\label{tab:incomplete-model-performance}
\begin{tabular}{llrrr}
\toprule
Model & Prompt & Accuracy (\%) & FPR (\%) & FNR (\%) \\
\midrule
GPT-5.5 Thinking High & Domain-Aware & $90.57 \pm 1.50$ & 3.95 & 9.38 \\
GPT-5.5 Thinking High & Domain-Anonymized & $93.57 \pm 1.37$ & 3.39 & 5.20 \\
Claude Sonnet 5 & Domain-Aware & $76.28 \pm 3.16$ & 16.96 & 15.44 \\
Claude Sonnet 5 & Domain-Anonymized & $81.71 \pm 3.51$ & 23.46 & 7.45 \\
Claude Opus 5 & Domain-Aware & $94.93 \pm 1.12$ & 4.26 & 1.82 \\
Claude Opus 5 & Domain-Anonymized & $94.66 \pm 1.16$ & 3.56 & 2.16 \\
\bottomrule
\end{tabular}
\end{table}

Available-case performance for the exploratory configurations is summarized in Table~\ref{tab:incomplete-model-performance}. GPT-5.5 Thinking High achieved higher available-case accuracy than GPT-5.5 Instant and GPT-5.5 Thinking Medium under both prompt conditions. Claude Opus 5 showed a larger advantage over Claude Sonnet 5. Claude Opus 5 achieved $94.93\%$ $(\pm 1.12\%)$ accuracy for Domain-Aware and $94.66\%$ $(\pm 1.16\%)$ for Domain-Anonymized prompts, compared with $76.28\%$ $(\pm 3.16\%)$ and $81.71\%$ $(\pm 3.51\%)$ for Claude Sonnet 5, respectively. Claude Opus 5 also had low FNRs of 1.82\% and 2.16\%, whereas Claude Sonnet 5 showed a different error profile, including a Domain-Anonymized FPR of 23.46\%. Claude Opus 5 showed higher classification accuracy than Claude Sonnet 5 among the available responses. Domain-Anonymized prompting improved classification accuracy for both Claude Sonnet 5 and GPT-5.5 Thinking High among the available responses.

On the 1,584 response keys completed by all three GPT-5.5 configurations, GPT-5.5 Instant, GPT-5.5 Thinking Medium, and GPT-5.5 Thinking High achieved combined accuracies of 91.45\%, 91.07\%, and 92.05\%, respectively. Under Domain-Anonymized prompting on the aligned subset, the corresponding values were 92.84\%, 93.08\%, and 93.55\%. GPT-5.5 Thinking High retained the highest accuracy in these aligned comparisons, but its advantage over GPT-5.5 Instant and GPT-5.5 Thinking Medium was small. Taken together with the substantially better performance of Claude Opus 5 than Claude Sonnet 5, the exploratory findings suggest that higher-capability or higher-reasoning configurations may be better able to classify structural identifiability. However, the exploratory configurations did not achieve complete benchmark coverage, and their missing responses may have been concentrated among more difficult scenarios.

\begin{figure}[H]
\centering
\includegraphics[width=0.9\linewidth]{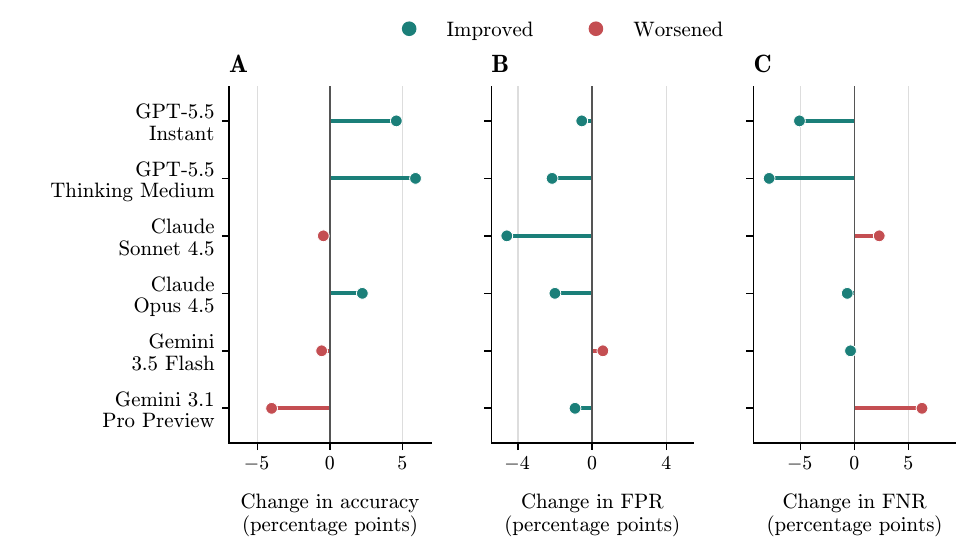}
\caption{\textbf{Performance shifts following Domain-Anonymization.} (A) Three-class accuracy, (B) False-positive rate (FPR), and (C) False-negative rate (FNR). Higher accuracy and lower error rates indicate improvement. Teal indicates improvement, red indicates worsening, and the vertical line at zero indicates no difference between prompt conditions.}
\label{fig:anonymization-effects}
\end{figure}

Anonymization improved classification performance for several configurations, with the clearest and most consistent gains observed for GPT-5.5 (Fig~\ref{fig:anonymization-effects}). Accuracy increased from 86.42\% to 90.99\% for GPT-5.5 Instant, from 85.13\% to 91.02\% for GPT-5.5 Thinking Medium, and from 49.01\% to 51.24\% for Claude Opus 4.5, while decreasing slightly for Claude Sonnet 4.5 (50.89\% versus 50.43\%). Anonymization reduced both FP and FN for the two GPT-5.5 configurations and Claude Opus 4.5. These results indicate that replacing model-specific state and parameter names with generic symbols can improve structural-identifiability classification in some LLM configurations. The effect was not universal, however, as accuracy decreased for Gemini 3.5 Flash and Gemini 3.1 Pro Preview. For Gemini 3.1 Pro Preview, accuracy decreased from 59.77\% to 55.75\%, and its FNR increased from 53.43\% to 59.66\% despite a decrease in FPR from 2.58\% to 1.65\%. 

The exploratory configurations showed a broadly similar but heterogeneous pattern. Domain-Anonymized prompting increased the available-case accuracy of GPT-5.5 Thinking High from $90.57\%$ $(\pm 1.50\%)$ to $93.57\%$ $(\pm 1.37\%)$ and reduced both its FPR (3.95\% to 3.39\%) and FNR (9.38\% to 5.20\%). Claude Sonnet 5 also showed higher accuracy after anonymization, increasing from $76.28\%$ $(\pm 3.16\%)$ to $81.71\%$ $(\pm 3.51\%)$; however, its lower FNR was accompanied by an increase in FPR from 16.96\% to 23.46\%. Claude Opus 5 changed little, with accuracy decreasing slightly from $94.93\%$ $(\pm 1.12\%)$ to $94.66\%$ $(\pm 1.16\%)$. Thus, the exploratory results provide additional evidence that anonymization can improve performance, particularly for the GPT configurations, but also show that its effect depends on the model and the type of classification error.

Because aggregate accuracy can conceal differences among identifiability classes and among scenarios of different complexity, the overall anonymization results alone do not show where these gains or losses occurred. We therefore examined performance separately by formal identifiability class and subsequently by the number of states, unknown parameters, observed outputs, and initial-condition setting.

\subsection*{Performance across formal identifiability classes}

The reference standard contained 477 globally identifiable, 25 locally identifiable, and 194 non-identifiable parameters per complete repeat. To examine differences in performance across identifiability classes,  Table~\ref{tab:formal-class-recall} presents class-specific recall for GPT-5.5 Instant, GPT-5.5 Thinking Medium, and GPT-5.5 Thinking High and for Claude Sonnet 5 and Claude Opus 5. The complete results for Claude Sonnet 4.5, Claude Opus 4.5, and Gemini configurations are provided in S6 Table.

\begin{table}[htbp]
\centering
\footnotesize
\caption{Class-specific recall for globally identifiable, locally identifiable, and non-identifiable parameters in selected primary and exploratory configurations. The values for GPT-5.5 Instant and GPT-5.5 Thinking Medium pool all ten complete repeats. The values for GPT-5.5 Thinking High and Claude 5 configurations are available-case estimates from successfully returned responses.}
\label{tab:formal-class-recall}
\begin{tabular}{llrrr}
\toprule
Model & Prompt & Global (\%) & Local (\%) & Non-identifiable (\%) \\
\midrule
GPT-5.5 Instant & Domain-Aware & 86.33 & 34.80 & 93.30 \\
GPT-5.5 Instant & Domain-Anonymized & 92.39 & 42.00 & 93.87 \\
GPT-5.5 Thinking Medium & Domain-Aware & 85.14 & 31.20 & 92.06 \\
GPT-5.5 Thinking Medium & Domain-Anonymized & 92.45 & 38.80 & 94.23 \\
GPT-5.5 Thinking High & Domain-Aware & 89.91 & 53.57 & 96.05 \\
GPT-5.5 Thinking High & Domain-Anonymized & 94.34 & 44.64 & 96.61 \\
Claude Sonnet 5 & Domain-Aware & 75.84 & 32.49 & 83.04 \\
Claude Sonnet 5 & Domain-Anonymized & 86.68 & 22.50 & 76.54 \\
Claude Opus 5 & Domain-Aware & 97.21 & 46.80 & 95.74 \\
Claude Opus 5 & Domain-Anonymized & 96.51 & 46.40 & 96.44 \\
\bottomrule
\end{tabular}
\end{table}

Under strict three-class scoring, global and non-identifiable parameters were classified substantially more accurately than locally identifiable parameters. Under Domain-Anonymized prompting, global recall was 92.39\% for GPT-5.5 Instant, 92.45\% for GPT-5.5 Thinking Medium, 94.34\% for GPT-5.5 Thinking High, 86.68\% for Claude Sonnet 5, and 96.51\% for Claude Opus 5. Non-identifiable recall ranged from 76.54\% for Claude Sonnet 5 to 96.61\% for GPT-5.5 Thinking High. Strict local recall remained lower than recall for the other classes. Its highest observed value was 53.57\% for GPT-5.5 Thinking High with Domain-Aware prompts, while Claude Opus 5 achieved 46.80\% with Domain-Aware and 46.40\% with Domain-Anonymized prompts.

Anonymization improved global and non-identifiable recall for all three GPT-5.5 configurations, but its effect on strict local recall was not consistent. Strict local recall increased from 34.80\% to 42.00\% for GPT-5.5 Instant and from 31.20\% to 38.80\% for GPT-5.5 Thinking Medium, but decreased from 53.57\% to 44.64\% for GPT-5.5 Thinking High. Claude Opus 5 maintained high global and non-identifiable recall under both prompt conditions, whereas Claude Sonnet 5 showed a larger increase in global recall but decreases in local and non-identifiable recall after anonymization. With only 25 retained local references, these estimates remain sensitive to the small size and composition of the local identifiable class.

\subsection*{Performance by scenario complexity}

\begin{figure}[H]
\centering
\includegraphics[width=0.9\linewidth]{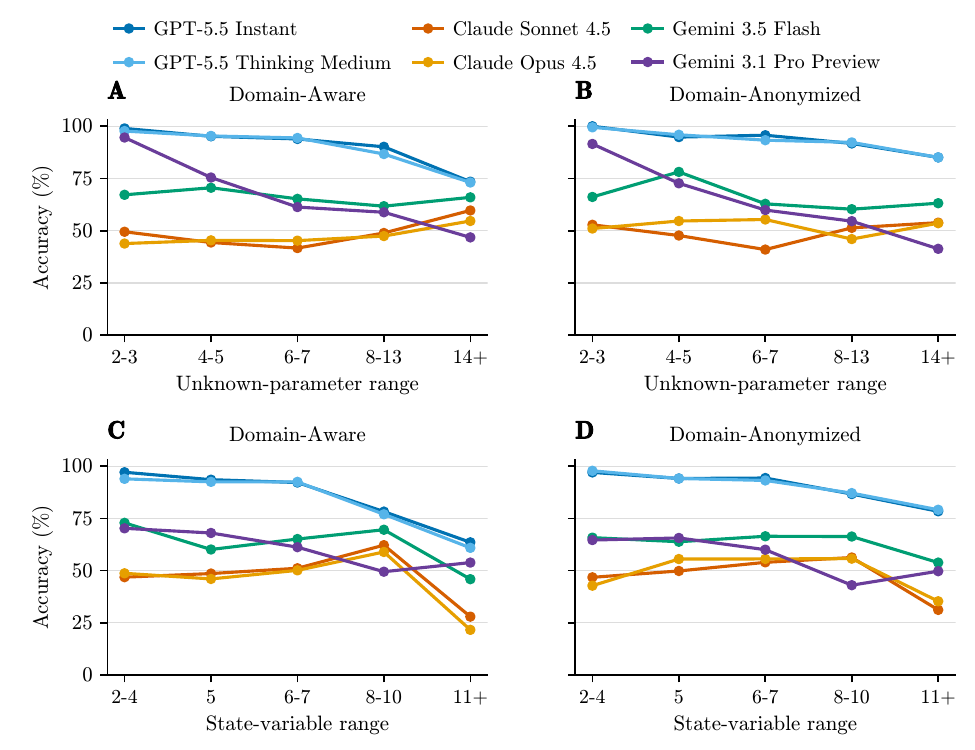}
\caption{\textbf{Model performance by scenario complexity and prompt condition.} Three-class parameter accuracy across five unknown-parameter ranges under (A) Domain-Aware and (B) Domain-Anonymized prompt and across five state-variable ranges under (C) Domain-Aware and (D) Domain-Anonymized prompt.}
\label{fig:scenario-complexity}
\end{figure}

The numbers of unknown parameters and state variables were each divided into five ranges selected to avoid sparsely populated exact-count categories (Fig~\ref{fig:scenario-complexity}, S7 Table). Under Domain-Aware prompting, accuracy declined substantially for the two GPT configurations as the number of unknown parameters increased. In the range with 14 or more unknown parameters, GPT-5.5 Instant accuracy increased from 73.46\% under Domain-Aware prompting to 85.09\% under Domain-Anonymized prompting, while GPT-5.5 Thinking Medium increased from 73.07\% to 85.09\%. The differences between prompt conditions were small in the lower parameter ranges and became more pronounced in the higher-complexity ranges. Claude and Gemini did not show the same consistent anonymization pattern, indicating that this effect was model-dependent.

A similar pattern was observed for the number of state variables. With at least 11 states, GPT-5.5 Instant accuracy increased from 63.53\% under Domain-Aware prompting to 78.38\% under Domain-Anonymized prompting, while GPT-5.5 Thinking Medium increased from 60.88\% to 79.12\%. Anonymization also improved the two GPT-5.5 configurations in the 8--10-state range, from 78.23\% to 86.65\% for GPT-5.5 Instant and from 76.95\% to 87.09\% for GPT-5.5 Thinking Medium. In contrast, little difference was observed in the lowest state ranges. Thus, the decline in accuracy of GPT-5.5 configurations with increasing scenario complexity was attenuated under Domain-Anonymized prompting, particularly in scenarios with at least 14 unknown parameters or at least 11 state variables.

The exploratory configurations showed a different complexity profile (S8 Table). Claude Opus 5 maintained high accuracy across all five unknown-parameter ranges, including 94.88\% with Domain-Aware and 93.93\% with Domain-Anonymized prompts in the range with 14 or more unknown parameters. It also retained approximately 87\% accuracy in scenarios with at least 11 states. For GPT-5.5 High, anonymization increased accuracy from 77.65\% to 87.32\% with 14 or more unknown parameters and from 78.85\% to 87.82\% with 8--10 states. Claude Sonnet 5 was less accurate and more variable than Claude Opus 5, but anonymization increased its accuracy from 67.87\% to 85.71\% in the highest unknown-parameter range and from 65.54\% to 81.00\% in the 8--10-state range. These available-case patterns suggest that the higher-capability Claude Opus 5 was more robust to scenario complexity, whereas GPT-5.5 High and Claude Sonnet 5 benefited from anonymization in several complex ranges.

Accuracy generally increased as more outputs were observed, and anonymization improved both GPT-5.5 configurations in all three output ranges (Table~\ref{tab:scenario-characteristics-accuracy}). The largest changes occurred with two outputs: GPT-5.5 Instant increased from 82.26\% to 89.15\%, and GPT-5.5 Thinking Medium increased from 78.87\% to 89.06\%. With at least three outputs, Domain-Anonymized accuracy reached 98.01\% for GPT-5.5 Instant and 97.30\% for GPT-5.5 Thinking Medium. The complete-configuration results for the other model families are provided in S9 Table.

Initial-condition effects also differed by prompt condition. In known and unknown settings, anonymization produced modest improvements for both GPT-5.5 configurations. In unknown settings, GPT-5.5 Instant increased from 86.29\% to 89.63\%, and GPT-5.5 Thinking Medium increased from 85.31\% to 89.31\%. The largest changes occurred in mixed settings, where Instant increased from 56.39\% to 77.38\%, and GPT-5.5 Thinking Medium increased from 49.02\% to 79.02\%. Because only four scenarios had mixed initial conditions, the magnitude of this contrast should be interpreted cautiously.

\begin{table}[htbp]
\centering
\footnotesize
\caption{Three-class accuracy (\%) by observed-output count, initial-condition setting, and prompt condition for GPT-5.5 Instant and GPT-5.5 Thinking Medium. All ten repeats are pooled.}
\label{tab:scenario-characteristics-accuracy}
\setlength{\tabcolsep}{4pt}
\begin{tabular}{lrlrr}
\toprule
Value or setting & Scenarios & Prompt condition
& \shortstack{GPT-5.5\\Instant}
& \shortstack{GPT-5.5\\Thinking Medium} \\
\midrule
\multicolumn{5}{l}{\textbf{A. Number of observed outputs}} \\
\addlinespace
1 & 40 & Domain-Aware & 80.00 & 80.00 \\
1 & 40 & Domain-Anonymized & 84.15 & 85.21 \\
\addlinespace
2 & 28 & Domain-Aware & 82.26 & 78.87 \\
2 & 28 & Domain-Anonymized & 89.15 & 89.06 \\
\addlinespace
3+ & 20 & Domain-Aware & 94.94 & 94.27 \\
3+ & 20 & Domain-Anonymized & 98.01 & 97.30 \\
\midrule
\multicolumn{5}{l}{\textbf{B. Initial-condition setting}} \\
\addlinespace
Known & 30 & Domain-Aware & 94.69 & 94.47 \\
Known & 30 & Domain-Anonymized & 97.06 & 97.28 \\
\addlinespace
Mixed & 4 & Domain-Aware & 56.39 & 49.02 \\
Mixed & 4 & Domain-Anonymized & 77.38 & 79.02 \\
\addlinespace
Unknown & 54 & Domain-Aware & 86.29 & 85.31 \\
Unknown & 54 & Domain-Anonymized & 89.63 & 89.31 \\
\bottomrule
\end{tabular}
\end{table}

Results from the exploratory configurations showed a similar relationship between output availability and classification accuracy, with higher accuracy in scenarios containing more observed outputs (S10 Table). Under Domain-Anonymized prompting, accuracy increased from 85.12\% with one output to 99.57\% with three or more outputs for GPT-5.5 Thinking High and from 92.70\% to 97.58\% for Claude Opus 5. Claude Sonnet 5 achieved 77.20\% with two outputs and 82.24\% with three or more outputs. Across initial-condition settings, Claude Opus 5 maintained high Domain-Anonymized accuracy for both known (96.57\%) and unknown (95.75\%) initial conditions.

\subsection*{Multivariable analysis of scenario complexity}

Because the scenario characteristics were interrelated, we used multivariable regression to examine the association of each characteristic with correct classification while accounting for the others (S11 Table). After adjustment, each additional unknown parameter was associated with lower odds of a correct classification (OR 0.881, 95\% CI 0.852--0.910), whereas each additional observed output was associated with higher odds (OR 1.666, 95\% CI 1.503--1.846). State count was not associated with correctness after adjustment (OR 0.970 per additional state, 95\% CI 0.912--1.033). Mixed initial-condition settings had lower adjusted odds than known settings, whereas the difference between unknown and known initial-condition settings was not statistically distinguishable. Interaction tests showed that the associations with unknown-parameter count and output count differed among LLM configurations (both $p<0.001$), while the association with state count did not ($p=0.114$).

\begin{figure}[H]
\centering
\includegraphics[width=0.9\linewidth]{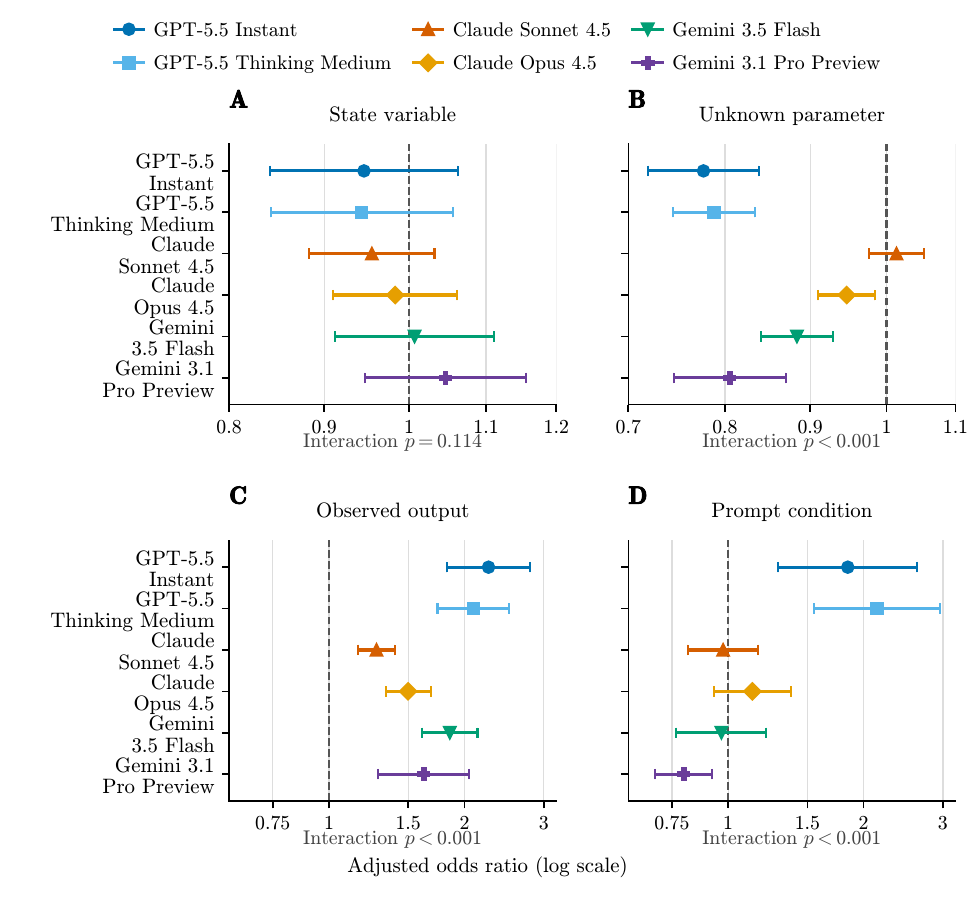}
\caption{\textbf{Adjusted associations by LLM configuration.}
Points and horizontal lines show odds ratios and 95\% confidence
intervals for (A) state count, (B) unknown-parameter count,
(C) observed-output count, and (D) Domain-Anonymized versus
Domain-Aware prompting. Count effects are per additional variable,
parameter, or output. The dashed line marks OR = 1.
Interaction $p$ values assess differences among configurations.
Estimates use scenario-clustered robust standard errors.}
\label{fig:adjusted-associations}
\end{figure}

Model-specific estimates from the interaction models further clarified these associations (Fig~\ref{fig:adjusted-associations}; S12 Table). Each additional unknown parameter was associated with lower odds of correctness for five of the six complete configurations, with ORs ranging from 0.777 for GPT Instant to 0.946 for Claude Opus 4.5; Claude Sonnet 4.5 was the exception (OR 1.014, 95\% CI 0.976--1.053). Each additional observed output was positively associated with correctness in all six configurations, although the magnitude ranged from OR 1.275 for Claude Sonnet 4.5 to OR 2.262 for GPT-5.5 Instant. State-count estimates were not statistically distinguishable from one for any configuration. The adjusted effect of anonymization also differed across configurations ($p<0.001$ for the prompt-by-LLM interaction). Anonymization was associated with higher odds of correctness for GPT-5.5 Instant (OR 1.844, 95\% CI 1.292--2.632) and GPT-5.5 Thinking Medium (OR 2.139, 95\% CI 1.550--2.952), but with lower odds for Gemini 3.1 Pro Preview (OR 0.797, 95\% CI 0.689--0.921). The corresponding estimates for Claude Sonnet 4.5, Claude Opus 4.5, and Gemini 3.5 Flash were not statistically distinguishable from one. Thus, both the complexity associations and the anonymization effect were model-dependent.

\section*{Discussion}
In this retrospective benchmark, we evaluated six complete LLM configurations across 88 epidemic-model scenarios using two prompt conditions and ten repeated runs. Four primary findings stand out. First, GPT-5.5 Instant and GPT-5.5 Thinking Medium achieved higher observed parameter-level accuracies than the other complete configurations, while their own accuracies were similar. Thinking High and Claude Opus 5 also showed comparatively high observed available-case performance, consistent with the possibility that greater reasoning effort or a higher-capability configuration may improve structural identifiability classification; complete benchmark coverage is needed to evaluate this possibility more firmly. Second, locally identifiable parameters remained the weakest class under strict scoring; however, the local reference set was small and concentrated. Third, name anonymization improved both complete GPT-5.5 configurations and Claude Opus 4.5, with particularly large GPT-5.5 configuration gains in scenarios containing at least 14 unknown parameters or at least 11 state variables, but the effect was not universal across model families. Fourth, a greater unknown-parameter count was associated with lower correctness and a greater number of observed output functions was associated with higher correctness after mutual adjustment; these are descriptive associations rather than evidence that the outputs are informationally equivalent or causally beneficial. These findings do not establish that the evaluated LLM proves structural identifiability. The output remains a classification hypothesis. Formal paper-based analysis or symbolic computation remains the evidentiary standard.

Preserved or improved performance after canonical symbolic relabeling indicates that the evaluated configurations did not rely exclusively on the original disease-specific state and parameter labels. This is consistent with broader evidence that mathematically equivalent symbolic reformulations can alter LLM reasoning performance \cite{mirzadeh2024gsmsymbolic}. However, the relabeling also changed tokenization and notational regularity and may have introduced index or ordering cues while preserving equation topology, observed outputs, and initial-condition assumptions. It therefore does not isolate terminology and does not exclude memorization of equation structures or source-study conclusions. The heterogeneous effects across model families further caution against interpreting canonical relabeling as a universal improvement.

The exploratory results showed higher observed performance for some higher-effort or higher-capability configurations. GPT-5.5 Thinking High had higher available-case accuracy than GPT-5.5 Instant and GPT-5.5 Thinking Medium, while Claude Opus 5 had higher observed accuracy than Claude Sonnet 5. These patterns are consistent with a possible performance benefit from greater reasoning effort or a higher model tier, although complete benchmark coverage is needed for a definitive comparison.

LLM performance varied with scenario complexity. In the stratified analysis, accuracy generally decreased as the number of states increased and increased as more outputs were observed, whereas the relationship with unknown-parameter count was not monotonic. After adjustment for overlapping scenario characteristics, additional unknown parameters were associated with lower odds of correct classification, while additional observed outputs were associated with higher odds. State count was no longer associated with correctness after adjustment, suggesting that its unadjusted relationship with accuracy may reflect other correlated complexity characteristics. The magnitude of these associations differed among LLM configurations. Because the benchmark characteristics were not independently varied, these results represent associations rather than causal effects.

Local identifiability remained the most challenging class under strict three-class scoring. The retained reference standard contained 25 local parameters. These results suggest a genuine model-dependent weakness in distinguishing finite ambiguity from global uniqueness and non-identifiability, although the small and concentrated local class remains a reason for cautious interpretation and for future benchmarks with more independently confirmed local references. The repeated-run analysis shows that consistency does not necessarily imply correctness: repeatedly returning the same classification can reflect a persistent error. These descriptive findings suggest that LLMs may generate preliminary classification hypotheses or flag cases for formal review, provided that each classification is verified against published analytical results or formal symbolic methods for the same model, observed outputs, unknown parameters, and initial-condition assumptions. High parameter-level accuracy and repeated agreement do not establish that every parameter has been correctly classified. LLM outputs should therefore not be used to certify the structural identifiability of a complete model or, without formal verification, to support mechanistic interpretations or intervention conclusions.

Several limitations should be considered. First, the reference standard combined classifications reported in the source papers with results obtained using \texttt{StructuralIdentifiability.jl}, and the Julia-derived classifications were not independently confirmed using a second symbolic implementation. Second, the benchmark included only scenarios for which the model specification and a directly comparable formal classification were available. The resulting selection and exclusion process may limit the generalizability of the findings to other epidemic models and observation settings. Third, the scenario characteristics used to represent complexity were correlated. Although the multivariable analysis adjusted for these characteristics simultaneously, residual confounding may remain, and the reported associations should not be interpreted causally. Fourth, training-data contamination cannot be ruled out because some source papers, equations, or identifiability conclusions may have appeared in model training corpora \cite{golchin2024timetravel}. Preserved or improved performance after canonical symbolic relabeling indicates only that the evaluated configurations did not rely exclusively on the original variable names; it does not exclude memorization of equation structures, source-study results, or published conclusions. Fifth, the gateway aliases were not provider-version-pinned, and provider defaults and any implicit caching were not fully observable. Finally, this benchmark evaluated structural rather than practical identifiability and did not address parameter estimation from finite or noisy data. The evaluation also focused on individual parameter classifications; identifiable parameter combinations require a separate analysis and were not included in the reported accuracy measures.

\section*{Conclusion}

In this retrospective benchmark, we systematically evaluated the capability of LLMs to accurately classify the structural identifiability of epidemic models. Among the six LLM configurations that achieved complete ten-run coverage, GPT-5.5 Instant and GPT-5.5 Thinking Medium demonstrated the highest observed agreement with the formal reference classifications, although their overall accuracies were similar. In our exploratory analyses, GPT-5.5 Thinking High and Claude Opus 5 exhibited comparatively high performance, supporting the hypothesis that increased reasoning effort or higher-tier model capabilities can enhance structural identifiability classification. Domain anonymization improved the GPT-5.5 configurations and attenuated their performance decline in several complex scenario groups; however, this benefit varied significantly across model families and could not be attributed solely to the removal of biological terminology. Even after we excluded scenarios containing unresolved, at-least-local references, strict local recall remained consistently lower than both global and non-identifiable recall. Following statistical adjustment for overlapping scenario characteristics, each additional unknown parameter was associated with reduced classification correctness, whereas an increased number of observed output functions correlated with higher correctness. The magnitude of these descriptive relationships differed depending on the specific LLM configuration evaluated. These findings suggest that LLMs may generate preliminary structural identifiability hypotheses or flag cases for formal review, provided their classifications are verified using published analytical results or formal symbolic methods under the same model specification. They should not be used to certify the structural identifiability of a complete model.

\section*{Data availability}
The frozen Domain-Aware and Domain-Anonymized prompts, scenario and parameter reference tables with provenance and uncertainty flags, and formal Julia scripts are publicly available at \url{https://github.com/geunsoojang/llm-structural-identifiability-benchmark}. Files are organized by benchmark, source paper, and prompt condition. The original LLM responses used in the reported analyses, together with response identifiers and coverage inventories, are archived on Zenodo at \url{https://doi.org/10.5281/zenodo.22715089}.

\section*{Supporting information}

\setcounter{figure}{0}
\renewcommand{\thefigure}{S\arabic{figure}}

\paragraph*{S1 Table.}
\label{S1_Table}
{\bf Benchmark scenario specifications and model equations.}

\par\addvspace{\baselineskip}
\noindent
\begin{minipage}{\linewidth}

\paragraph*{S2 Table.}
\label{S2_Table}
{\bf Response completion  rates and API configurations for the standardized benchmark and subsequent follow-up experiments.} Every listed experiment was executed via the API interface. The six configurations with complete full-benchmark coverage formed the primary comparison. Incomplete configurations were analyzed separately.
\begin{center}
\scriptsize
\begin{tabular}{@{}p{0.24\linewidth}p{0.27\linewidth}p{0.18\linewidth}p{0.09\linewidth}p{0.12\linewidth}@{}}
\toprule
Model configuration & Gateway identifier & Completed/scheduled & Coverage (\%) & Analysis set \\
\midrule
GPT-5.5 Instant & \url{openai/gpt5_5} & 1,760/1,760 & 100.00 & Primary \\
GPT-5.5 Thinking Medium & \url{openai/gpt5_5_thinking} & 1,760/1,760 & 100.00 & Primary \\
Claude Sonnet 4.5 & \url{aws/claude4_5_sonnet} & 1,760/1,760 & 100.00 & Primary \\
Claude Opus 4.5 & \url{aws/claude4_5_opus} & 1,760/1,760 & 100.00 & Primary \\
Gemini 3.5 Flash & \url{gcp-deepmind/geminiflash3_5} & 1,760/1,760 & 100.00 & Primary \\
Gemini 3.1 Pro Preview & \url{gcp-deepmind/geminipro3_1} & 1,760/1,760 & 100.00 & Primary \\
GPT-5.5 Thinking High & \url{openai/gpt5_5_thinking} & 1,584/1,760 & 90.00 & Exploratory \\
Claude Sonnet 5 & \url{aws/claude5_sonnet} & 1,498/1,760 & 85.11 & Exploratory \\
Claude Opus 5 & \url{aws/claude5_opus} & 1,732/1,760 & 98.41 & Exploratory \\
\bottomrule
\end{tabular}
\end{center}

\end{minipage}
\par\addvspace{\baselineskip}

\par\addvspace{\baselineskip}
\noindent
\begin{minipage}{\linewidth}

\paragraph*{S3 Table.}
\label{S3_Table}
{\bf Group definitions and scenario counts used in the descriptive complexity analyses.} Each of the 88 benchmark scenarios contributed once to the distribution for each characteristic. Percentages are calculated within each characteristic and may not sum to exactly 100\% because of rounding.
\begin{center}
\small
\begin{tabular}{llrr}
\toprule
Characteristic & Group & Scenarios & Percentage (\%) \\
\midrule
Unknown parameters & 2--3 & 16 & 18.18 \\
 & 4--5 & 24 & 27.27 \\
 & 6--7 & 13 & 14.77 \\
 & 8--13 & 22 & 25.00 \\
 & 14+ & 13 & 14.77 \\
\addlinespace
State variables & 2--4 & 24 & 27.27 \\
 & 5 & 22 & 25.00 \\
 & 6--7 & 21 & 23.86 \\
 & 8--10 & 12 & 13.64 \\
 & 11+ & 9 & 10.23 \\
\addlinespace
Observed outputs & 1 & 40 & 45.45 \\
 & 2 & 28 & 31.82 \\
 & 3+ & 20 & 22.73 \\
\bottomrule
\end{tabular}
\end{center}

\end{minipage}
\par\addvspace{\baselineskip}

\par\addvspace{\baselineskip}
\noindent
\begin{minipage}{\linewidth}

\paragraph*{S4 Table.}
\label{S4_Table}
{\bf Detailed three-class accuracy and binary identifiability errors for configurations with complete coverage.} Accuracy is presented as mean $\pm$ standard deviation across ten separately submitted runs. TP, FP, FN, and TN are pooled counts across runs. Global and local classes are grouped as identifiable for the binary analysis; accuracy, FPR, and FNR are percentages. Abstain denotes missing, conflicting, or unparseable parameter classifications, which were counted as incorrect in three-class accuracy and excluded from binary FPR and FNR calculations.
\begin{center}
\scriptsize
\resizebox{\textwidth}{!}{%
\begin{tabular}{llrrrrrrrr}
\toprule
Model & Prompt & Accuracy (\%) & TP & FP & FN & TN & FPR & FNR & Abstain \\
\midrule
GPT-5.5 Instant & Domain-Aware & 86.42 $\pm$ 2.05 & 4406 & 130 & 613 & 1810 & 6.70 & 12.21 & 1 \\
GPT-5.5 Instant & Domain-Anonymized & 90.99 $\pm$ 1.80 & 4663 & 119 & 357 & 1821 & 6.13 & 7.11 & 0 \\
GPT-5.5 Thinking Medium & Domain-Aware & 85.13 $\pm$ 1.56 & 4321 & 154 & 698 & 1786 & 7.94 & 13.91 & 1 \\
GPT-5.5 Thinking Medium & Domain-Anonymized & 91.02 $\pm$ 2.16 & 4719 & 112 & 301 & 1828 & 5.77 & 6.00 & 0 \\
Claude Sonnet 4.5 & Domain-Aware & 50.89 $\pm$ 0.93 & 2173 & 374 & 2827 & 1533 & 19.61 & 56.54 & 53 \\
Claude Sonnet 4.5 & Domain-Anonymized & 50.43 $\pm$ 0.85 & 2051 & 287 & 2929 & 1625 & 15.01 & 58.82 & 68 \\
Claude Opus 4.5 & Domain-Aware & 49.01 $\pm$ 0.45 & 1883 & 252 & 3119 & 1628 & 13.40 & 62.36 & 78 \\
Claude Opus 4.5 & Domain-Anonymized & 51.24 $\pm$ 0.24 & 1924 & 221 & 3096 & 1719 & 11.39 & 61.67 & 0 \\
Gemini 3.5 Flash & Domain-Aware & 65.22 $\pm$ 1.65 & 3012 & 319 & 2008 & 1621 & 16.44 & 40.00 & 0 \\
Gemini 3.5 Flash & Domain-Anonymized & 64.64 $\pm$ 2.11 & 3031 & 330 & 1989 & 1610 & 17.01 & 39.62 & 0 \\
Gemini 3.1 Pro Preview & Domain-Aware & 59.77 $\pm$ 1.68 & 2338 & 50 & 2682 & 1888 & 2.58 & 53.43 & 2 \\
Gemini 3.1 Pro Preview & Domain-Anonymized & 55.75 $\pm$ 2.21 & 2025 & 32 & 2995 & 1908 & 1.65 & 59.66 & 0 \\
\bottomrule
\end{tabular}%
}
\end{center}

\end{minipage}
\par\addvspace{\baselineskip}

\par\addvspace{\baselineskip}
\noindent
\begin{minipage}{\linewidth}

\paragraph*{S5 Table.}
\label{S5_Table}
{\bf Repeated-run classification stability.} Values are $n/696$ (\%). Consistent valid requires the same valid predicted class on all ten submissions. Correct on all ten requires agreement with the formal reference on every submission; consistently wrong requires the same incorrect valid class on every submission.
\begin{center}
\scriptsize
\setlength{\tabcolsep}{3pt}
\begin{tabular}{llrrr}
\toprule
Configuration & Prompt
& \shortstack{Consistent valid\\$n/696$ (\%)}
& \shortstack{Correct on all ten\\$n/696$ (\%)}
& \shortstack{Consistently wrong\\$n/696$ (\%)} \\
\midrule
GPT-5.5 Instant & Domain-Aware
& 552/696 (79.31) & 506/696 (72.70) & 46/696 (6.61) \\
& Domain-Anonymized
& 561/696 (80.60) & 549/696 (78.88) & 12/696 (1.72) \\
\addlinespace
GPT-5.5 Thinking Medium & Domain-Aware
& 550/696 (79.02) & 499/696 (71.70) & 51/696 (7.33) \\
& Domain-Anonymized
& 582/696 (83.62) & 560/696 (80.46) & 22/696 (3.16) \\
\addlinespace
Claude Sonnet 4.5 & Domain-Aware
& 597/696 (85.78) & 312/696 (44.83) & 285/696 (40.95) \\
& Domain-Anonymized
& 616/696 (88.51) & 315/696 (45.26) & 301/696 (43.25) \\
\addlinespace
Claude Opus 4.5 & Domain-Aware
& 661/696 (94.97) & 330/696 (47.41) & 331/696 (47.56) \\
& Domain-Anonymized
& 671/696 (96.41) & 346/696 (49.71) & 325/696 (46.70) \\
\addlinespace
Gemini 3.5 Flash & Domain-Aware
& 494/696 (70.98) & 348/696 (50.00) & 146/696 (20.98) \\
& Domain-Anonymized
& 514/696 (73.85) & 360/696 (51.72) & 154/696 (22.13) \\
\addlinespace
Gemini 3.1 Pro Preview & Domain-Aware
& 489/696 (70.26) & 323/696 (46.41) & 166/696 (23.85) \\
& Domain-Anonymized
& 465/696 (66.81) & 291/696 (41.81) & 174/696 (25.00) \\
\bottomrule
\end{tabular}
\end{center}

\end{minipage}
\par\addvspace{\baselineskip}

\par\addvspace{\baselineskip}
\noindent
\begin{minipage}{\linewidth}

\paragraph*{S6 Table.}
\label{S6_Table}
{\bf Class-specific recall for complete Claude Sonnet 4.5, Claude Opus 4.5, Gemini 3.5 Flash, and Gemini Pro Preview.} Recall is the percentage of formal-reference parameters in each class that were correctly assigned to that class. Each prompt condition pools ten complete repeats.
\begin{center}
\small
\begin{tabular}{llrrr}
\toprule
Model & Prompt & Global (\%) & Local (\%) & Non-identifiable (\%) \\
\midrule
Claude Sonnet 4.5 & Domain-Aware & 42.12 & 0.00 & 79.02 \\
Claude Sonnet 4.5 & Domain-Anonymized & 39.52 & 0.00 & 83.76 \\
Claude Opus 4.5 & Domain-Aware & 37.38 & 0.00 & 83.92 \\
Claude Opus 4.5 & Domain-Anonymized & 38.72 & 0.00 & 88.61 \\
Gemini 3.5 Flash & Domain-Aware & 61.15 & 0.40 & 83.56 \\
Gemini 3.5 Flash & Domain-Anonymized & 60.57 & 0.00 & 82.99 \\
Gemini 3.1 Pro Preview & Domain-Aware & 46.77 & 16.40 & 97.32 \\
Gemini 3.1 Pro Preview & Domain-Anonymized & 40.67 & 12.80 & 98.35 \\
\bottomrule
\end{tabular}
\end{center}

\end{minipage}
\par\addvspace{\baselineskip}

\par\addvspace{\baselineskip}
\noindent
\begin{minipage}{\linewidth}

\paragraph*{S7 Table.}
\label{S7_Table}
{\bf Three-class accuracy by unknown-parameter count, state-variable count, and prompt condition for the six complete configurations.} The accuracy (\%) pooled across ten runs of the final 88-scenario benchmark. Each range is shown separately for Domain-Aware and Domain-Anonymized prompts. Scenarios denotes the number of distinct scenarios in each range, not the number of repeated responses.
\begin{center}
\small
\setlength{\tabcolsep}{5pt}
\resizebox{\linewidth}{!}{%
\begin{tabular}{lrlcccccc}
\toprule
Range & Scenarios & Prompt condition & \shortstack{GPT-5.5\\Instant} & \shortstack{GPT-5.5\\Thinking Medium} & \shortstack{Claude\\Sonnet 4.5} & \shortstack{Claude\\Opus 4.5} & \shortstack{Gemini\\3.5 Flash} & \shortstack{Gemini 3.1\\Pro Preview} \\
\midrule
\multicolumn{9}{l}{\textbf{A. Number of unknown parameters}} \\
\addlinespace
2--3 & 16 & Domain-Aware & 98.97 & 97.69 & 49.49 & 43.85 & 67.18 & 94.62 \\
2--3 & 16 & Domain-Anonymized & 100.00 & 99.49 & 52.82 & 51.03 & 66.15 & 91.54 \\
\addlinespace
4--5 & 24 & Domain-Aware & 95.19 & 95.37 & 44.35 & 45.37 & 70.56 & 75.46 \\
4--5 & 24 & Domain-Anonymized & 94.81 & 95.93 & 47.69 & 54.63 & 78.15 & 72.69 \\
\addlinespace
6--7 & 13 & Domain-Aware & 93.93 & 94.40 & 41.67 & 45.24 & 65.24 & 61.31 \\
6--7 & 13 & Domain-Anonymized & 95.71 & 93.33 & 40.95 & 55.36 & 62.86 & 59.88 \\
\addlinespace
8--13 & 22 & Domain-Aware & 90.17 & 86.71 & 48.90 & 47.43 & 61.73 & 58.82 \\
8--13 & 22 & Domain-Anonymized & 91.77 & 92.28 & 51.35 & 45.99 & 60.30 & 54.56 \\
\addlinespace
14+ & 13 & Domain-Aware & 73.46 & 73.07 & 59.69 & 54.65 & 65.96 & 46.80 \\
14+ & 13 & Domain-Anonymized & 85.09 & 85.09 & 53.86 & 53.60 & 63.16 & 41.32 \\
\addlinespace
\midrule
\multicolumn{9}{l}{\textbf{B. Number of state variables}} \\
\addlinespace
2--4 & 24 & Domain-Aware & 97.08 & 93.92 & 46.85 & 48.69 & 72.85 & 70.23 \\
2--4 & 24 & Domain-Anonymized & 97.00 & 97.77 & 46.77 & 42.77 & 65.77 & 64.62 \\
\addlinespace
5 & 22 & Domain-Aware & 93.52 & 92.48 & 48.57 & 46.00 & 60.10 & 68.00 \\
5 & 22 & Domain-Anonymized & 94.10 & 94.10 & 49.81 & 55.52 & 63.81 & 65.62 \\
\addlinespace
6--7 & 21 & Domain-Aware & 92.16 & 92.47 & 51.11 & 50.11 & 65.11 & 61.21 \\
6--7 & 21 & Domain-Anonymized & 94.32 & 93.16 & 53.95 & 55.58 & 66.42 & 60.00 \\
\addlinespace
8--10 & 12 & Domain-Aware & 78.23 & 76.95 & 62.17 & 58.92 & 69.56 & 49.46 \\
8--10 & 12 & Domain-Anonymized & 86.65 & 87.09 & 56.26 & 55.71 & 66.31 & 43.00 \\
\addlinespace
11+ & 9 & Domain-Aware & 63.53 & 60.88 & 27.94 & 21.62 & 45.88 & 53.82 \\
11+ & 9 & Domain-Anonymized & 78.38 & 79.12 & 31.18 & 35.29 & 53.82 & 49.71 \\
\addlinespace
\bottomrule
\end{tabular}%
}
\end{center}

\end{minipage}
\par\addvspace{\baselineskip}

\par\addvspace{\baselineskip}
\noindent
\begin{minipage}{\linewidth}

\paragraph*{S8 Table.}
\label{S8_Table}
{\bf Exploratory accuracy by unknown-parameter and state-variable ranges.}
The accuracy (\%) for successfully returned responses from GPT-5.5 Thinking High, Claude Sonnet 5, and Claude Opus 5. Each range is shown separately for Domain-Aware and Domain-Anonymized prompts.

\begin{center}
\footnotesize
\setlength{\tabcolsep}{4pt}
\begin{tabular}{llccc}
\toprule
Range & Prompt condition & \shortstack{GPT-5.5\\Thinking High} & \shortstack{Claude\\Sonnet 5} & \shortstack{Claude\\Opus 5} \\
\midrule
\multicolumn{5}{l}{\textbf{A. Number of unknown parameters}} \\
\addlinespace
2--3 & Domain-Aware & 98.95 & 99.23 & 98.72 \\
2--3 & Domain-Anonymized & 98.96 & 99.74 & 98.72 \\
\addlinespace
4--5 & Domain-Aware & 95.74 & 91.75 & 93.98 \\
4--5 & Domain-Anonymized & 94.86 & 88.15 & 93.98 \\
\addlinespace
6--7 & Domain-Aware & 99.61 & 85.13 & 97.22 \\
6--7 & Domain-Anonymized & 98.03 & 87.36 & 97.36 \\
\addlinespace
8--13 & Domain-Aware & 94.07 & 67.59 & 93.92 \\
8--13 & Domain-Anonymized & 95.35 & 68.63 & 93.96 \\
\addlinespace
14+ & Domain-Aware & 77.65 & 67.87 & 94.88 \\
14+ & Domain-Anonymized & 87.32 & 85.71 & 93.93 \\
\midrule
\multicolumn{5}{l}{\textbf{B. Number of state variables}} \\
\addlinespace
2--4 & Domain-Aware & 98.13 & 79.40 & 99.49 \\
2--4 & Domain-Anonymized & 99.07 & 81.30 & 98.77 \\
\addlinespace
5 & Domain-Aware & 97.07 & 79.86 & 94.38 \\
5 & Domain-Anonymized & 96.70 & 76.40 & 96.00 \\
\addlinespace
6--7 & Domain-Aware & 96.01 & 80.42 & 95.17 \\
6--7 & Domain-Anonymized & 95.98 & 83.79 & 94.98 \\
\addlinespace
8--10 & Domain-Aware & 78.85 & 65.54 & 94.99 \\
8--10 & Domain-Anonymized & 87.82 & 81.00 & 93.62 \\
\addlinespace
11+ & Domain-Aware & 78.84 & 78.37 & 87.06 \\
11+ & Domain-Anonymized & 83.61 & 84.75 & 87.21 \\
\bottomrule
\end{tabular}
\end{center}

\end{minipage}
\par\addvspace{\baselineskip}

\par\addvspace{\baselineskip}
\noindent
\begin{minipage}{\linewidth}

\paragraph*{S9 Table.}
\label{S9_Table}
{\bf Three-class accuracy (\%) by observed outputs count, initial condition setting, and prompt condition for the Claude and Gemini configurations.}

\begin{center}
\footnotesize
\setlength{\tabcolsep}{4pt}
\begin{tabular}{lrlcccc}
\toprule
Value or setting & Scenarios & Prompt condition
& \shortstack{Claude\\Sonnet 4.5}
& \shortstack{Claude\\Opus 4.5}
& \shortstack{Gemini\\3.5 Flash}
& \shortstack{Gemini 3.1\\Pro Preview} \\
\midrule
\multicolumn{7}{l}{\textbf{A. Number of observed outputs}} \\
\addlinespace
1 & 40 & Domain-Aware & 49.08 & 47.60 & 59.45 & 61.38 \\
1 & 40 & Domain-Anonymized & 49.68 & 54.88 & 60.92 & 58.53 \\
\addlinespace
2 & 28 & Domain-Aware & 40.75 & 33.54 & 53.07 & 57.26 \\
2 & 28 & Domain-Anonymized & 44.06 & 39.62 & 54.20 & 55.38 \\
\addlinespace
3+ & 20 & Domain-Aware & 60.41 & 62.43 & 79.55 & 60.45 \\
3+ & 20 & Domain-Anonymized & 56.10 & 57.49 & 75.96 & 53.78 \\
\midrule
\multicolumn{7}{l}{\textbf{B. Initial-condition setting}} \\
\addlinespace
Known & 30 & Domain-Aware & 41.84 & 38.95 & 68.82 & 56.89 \\
Known & 30 & Domain-Anonymized & 39.17 & 40.44 & 71.01 & 48.03 \\
\addlinespace
Mixed & 4 & Domain-Aware & 37.70 & 24.59 & 51.80 & 32.95 \\
Mixed & 4 & Domain-Anonymized & 23.61 & 32.62 & 38.03 & 26.89 \\
\addlinespace
Unknown & 54 & Domain-Aware & 57.94 & 58.30 & 65.21 & 65.41 \\
Unknown & 54 & Domain-Anonymized & 60.76 & 60.07 & 65.06 & 64.40 \\
\bottomrule
\end{tabular}
\end{center}

\end{minipage}
\par\addvspace{\baselineskip}

\par\addvspace{\baselineskip}
\noindent
\begin{minipage}{\linewidth}

\paragraph*{S10 Table.}
\label{S10_Table}
{\bf Exploratory accuracy by observed-output count, initial-condition setting, and prompt condition.}

\begin{center}
\footnotesize
\setlength{\tabcolsep}{4pt}
\begin{tabular}{lrlccc}
\toprule
Value or setting & Scenarios & Prompt condition
& \shortstack{GPT-5.5\\Thinking High}
& \shortstack{Claude\\Sonnet 5}
& \shortstack{Claude\\Opus 5} \\
\midrule
\multicolumn{6}{l}{\textbf{A. Number of observed outputs}} \\
\addlinespace
1 & 40 & Domain-Aware & 81.49 & 78.27 & 92.39 \\
1 & 40 & Domain-Anonymized & 85.12 & 84.40 & 92.70 \\
\addlinespace
2 & 28 & Domain-Aware & 92.41 & 74.80 & 92.85 \\
2 & 28 & Domain-Anonymized & 95.74 & 77.20 & 92.98 \\
\addlinespace
3+ & 20 & Domain-Aware & 97.60 & 75.44 & 98.62 \\
3+ & 20 & Domain-Anonymized & 99.57 & 82.24 & 97.58 \\
\midrule
\multicolumn{6}{l}{\textbf{B. Initial-condition setting}} \\
\addlinespace
Known & 30 & Domain-Aware & 95.20 & 75.99 & 96.66 \\
Known & 30 & Domain-Anonymized & 96.81 & 80.24 & 96.57 \\
\addlinespace
Mixed & 4 & Domain-Aware & 63.05 & 51.52 & 84.53 \\
Mixed & 4 & Domain-Anonymized & 84.85 & 62.69 & 78.44 \\
\addlinespace
Unknown & 54 & Domain-Aware & 89.74 & 80.73 & 95.43 \\
Unknown & 54 & Domain-Anonymized & 92.34 & 84.75 & 95.75 \\
\bottomrule
\end{tabular}
\end{center}

\end{minipage}
\par\addvspace{\baselineskip}

\par\addvspace{\baselineskip}
\noindent
\begin{minipage}{\linewidth}

\paragraph*{S11 Table.}
\label{S11_Table}
{\bf Adjusted associations between scenario characteristics and correct three-class classification.} Odds ratios are from the parameter-level logistic regression with scenario-clustered standard errors and adjustment for LLM configuration, formal class, prompt condition, and initial-condition setting.

\begin{center}
\footnotesize
\begin{tabular}{lccc}
\toprule
Term & Odds ratio & 95\% CI & $p$ value \\
\midrule
One additional state variable & 0.970 & 0.912--1.033 & 0.342 \\
One additional unknown parameter & 0.881 & 0.852--0.910 & $<0.001$ \\
One additional observed output & 1.666 & 1.503--1.846 & $<0.001$ \\
Domain-Anonymized versus Domain-Aware & 1.093 & 0.961--1.243 & 0.175 \\
Mixed versus known initial conditions & 0.451 & 0.259--0.785 & 0.005 \\
Unknown versus known initial conditions & 0.998 & 0.739--1.348 & 0.990 \\
\bottomrule
\end{tabular}
\end{center}

\end{minipage}
\par\addvspace{\baselineskip}

\par\addvspace{\baselineskip}
\noindent
\begin{minipage}{\linewidth}

\paragraph*{S12 Table.}
\label{S12_Table}
{\bf Model-specific adjusted associations with correct three-class classification for the six complete configurations.} Complexity estimates were obtained from the complexity-by-LLM interaction model, and anonymization estimates were obtained from the prompt-by-LLM interaction model. Odds ratios for state variables, unknown parameters, and observed outputs correspond to an increase of one. Models adjusted for the scenario characteristics, prompt condition, initial-condition setting, formal identifiability class, and LLM configuration. Standard errors were clustered by scenario.
\begin{center}
\scriptsize
\begin{tabular}{llrrr}
\toprule
Model & Term & Odds ratio & 95\% CI & $p$ value \\
\midrule
GPT-5.5 Instant & State variables (+1) & 0.946 & 0.842--1.062 & 0.344 \\
 & Unknown parameters (+1) & 0.777 & 0.719--0.838 & $<0.001$ \\
 & Observed outputs (+1) & 2.262 & 1.829--2.799 & $<0.001$ \\
 & Domain-Anonymized versus Domain-Aware & 1.844 & 1.292--2.632 & $<0.001$ \\
\addlinespace
GPT-5.5 Thinking Medium & State variables (+1) & 0.943 & 0.842--1.055 & 0.306 \\
 & Unknown parameters (+1) & 0.788 & 0.745--0.834 & $<0.001$ \\
 & Observed outputs (+1) & 2.093 & 1.742--2.514 & $<0.001$ \\
 & Domain-Anonymized versus Domain-Aware & 2.139 & 1.550--2.952 & $<0.001$ \\
\addlinespace
Claude Sonnet 4.5 & State variables (+1) & 0.955 & 0.883--1.032 & 0.243 \\
 & Unknown parameters (+1) & 1.014 & 0.976--1.053 & 0.476 \\
 & Observed outputs (+1) & 1.275 & 1.161--1.400 & $<0.001$ \\
 & Domain-Anonymized versus Domain-Aware & 0.975 & 0.814--1.167 & 0.781 \\
\addlinespace
Claude Opus 4.5 & State variables (+1) & 0.983 & 0.910--1.061 & 0.659 \\
 & Unknown parameters (+1) & 0.946 & 0.910--0.984 & 0.006 \\
 & Observed outputs (+1) & 1.499 & 1.336--1.682 & $<0.001$ \\
 & Domain-Anonymized versus Domain-Aware & 1.132 & 0.931--1.377 & 0.214 \\
\addlinespace
Gemini 3.5 Flash & State variables (+1) & 1.007 & 0.912--1.111 & 0.896 \\
 & Unknown parameters (+1) & 0.884 & 0.841--0.929 & $<0.001$ \\
 & Observed outputs (+1) & 1.856 & 1.611--2.138 & $<0.001$ \\
 & Domain-Anonymized versus Domain-Aware & 0.966 & 0.767--1.216 & 0.768 \\
\addlinespace
Gemini 3.1 Pro Preview & State variables (+1) & 1.046 & 0.946--1.156 & 0.378 \\
 & Unknown parameters (+1) & 0.805 & 0.745--0.870 & $<0.001$ \\
 & Observed outputs (+1) & 1.624 & 1.286--2.049 & $<0.001$ \\
 & Domain-Anonymized versus Domain-Aware & 0.797 & 0.689--0.921 & 0.002 \\
\bottomrule
\end{tabular}
\end{center}

\end{minipage}
\par\addvspace{\baselineskip}

\bibliography{references}

\end{document}